\documentclass[twocolumn]{aastex62}

\newcommand{\XMM}{\textit{XMM-Newton}}
\newcommand{\psr}{PSR~J0952$-$0607}
\newcommand{\Msun}{\ensuremath{M_{\rm Sun}}}
\newcommand{\DM}{\rm DM}
\newcommand{\NH}{\ensuremath{N_{\rm H}}}
\newcommand{\Teff}{\ensuremath{T_{\rm eff}}}
\newcommand{\Tc}{\ensuremath{T_{\rm c}}}
\newcommand{\nus}{\nu_{\rm s}}
\newcommand{\nudot}{\dot{\nu}_{\rm s}}
\newcommand{\tgw}{\ensuremath{t_{\rm gw}}}
\newcommand{\tvisc}{\ensuremath{t_{\rm visc}}}

\received{2019 May 28}
\revised{2019 July 19}
\accepted{2019 July 23}
\submitjournal{ApJ}

\shorttitle{\XMM\ detection of \psr}
\shortauthors{Ho, Heinke, and Chugunov}

\begin{document}

\title{\XMM\ detection and spectrum of the second fastest spinning pulsar \psr}

\correspondingauthor{Wynn Ho}
\email{wynnho@slac.stanford.edu}

\author[0000-0002-6089-6836]{Wynn C.G. Ho}
\affil{Department of Physics and Astronomy, Haverford College, 370 Lancaster Avenue, Haverford, PA 19041, USA}
\affil{Mathematical Sciences, Physics and Astronomy, and STAG Research Centre, \\ University of Southampton, Southampton SO17 1BJ, United Kingdom}

\author[0000-0003-3944-6109]{Craig O. Heinke}
\affiliation{Department of Physics, University of Alberta, CCIS 4-183, Edmonton, AB T6G 2E1, Canada}

\author[0000-0003-1333-6139]{Andrey I. Chugunov}
\affiliation{Ioffe Institute, 26 Politekhnicheskaya, St Petersburg 194021, Russia}

\begin{abstract}
With a spin frequency of 707~Hz,
PSR~J0952$-$0607 is the second fastest spinning pulsar known.
It was discovered in radio by LOFAR in 2017 at an estimated distance
of either 0.97 or 1.74~kpc and has a low-mass companion with a 6.42~hr
orbital period.
We report discovery of the X-ray counterpart of \psr\ using \XMM.
The X-ray spectra can be well-fit by a single power law model
($\Gamma\approx2.5$) or by a thermal plus power law model
($k\Teff\approx40\mbox{ eV}$ and $\Gamma\approx1.4$).
We do not detect evidence of variability, such as that due to
orbital modulation from pulsar wind and companion star interaction.
Because of its fast spin rate, \psr\ is a crucial source for understanding
the r-mode instability, which can be an effective mechanism for producing
gravitational waves.
Using the high end of our measured surface temperature, we infer a
neutron star core temperature of $\sim10^7\mbox{ K}$, which places
\psr\ within the window for the r-mode to be unstable unless an
effect such as superfluid mutual friction damps the fluid oscillation.
The measured luminosity limits the dimensionless r-mode amplitude to be
less than $\sim 1\times 10^{-9}$.
\end{abstract}


\section{Introduction} \label{sec:intro}

Despite sensitivity to and searches for pulsars with even higher spin rates,
the fastest pulsars known to date among non-accreting pulsars are
PSR~J1748$-$2446ad (with a spin rate $\nus=716$~Hz; \citealt{hesselsetal06}),
\psr\ ($\nus=707$~Hz; \citealt{bassaetal17}),
and PSR~B1937+21 ($\nus=642$~Hz; \citealt{backeretal82})
and among accreting pulsars are 4U~1608$-$52 ($\nus=620$~Hz),
SAX~J1750.8$-$2900 ($\nus=601$~Hz), and IGR~00291+5934 ($\nus=599$~Hz).
The observed spin rates are well below the theoretical limit
of $\sim$2000~Hz \citep{cooketal94,haenseletal99}.
This suggests a mechanism which prevents fast rotation
\citep{chakrabartyetal03,chakrabarty08,papittoetal14,patrunoetal17,gittinsandersson19},
such as mechanisms associated with gravitational wave emission since
their torques depend strongly on spin rate.

One particular mechanism that is of great interest is that associated
with the r-mode fluid oscillation, because it can be a strong source
of gravitational waves \citep{andersson98,friedmanmorsink98})
via the Chandrasekhar-Friedman-Schutz (CFS) instability
(\citealt{chandrasekhar70,friedmanschutz78}; see also \citealt{chugunov17}).
The strength of the r-mode instability is characterized by the balance
between the timescale of mode growth by gravitational wave emission
$\tgw$ (which depends strongly on spin frequency, i.e.,
$\tgw\propto\nus^{-6}$) and timescale of viscous damping
(which is temperature-dependent; e.g., $\tvisc\propto T^2$;
\citealt{anderssonkokkotas01}).
However, our theoretical picture of r-modes is severely problematic
(see, e.g., \citealt{hoetal11,haskelletal12,chugunovetal17}).
Thus it is vital to identify fast and hot neutron stars which can
be used to better understand the nature of the r-mode mechanism.
Here we report the X-ray detection of the second fastest pulsar, \psr,
which provides possibly the strongest r-mode constraints
for millisecond pulsars.

\psr\ was discovered at radio frequencies using LOFAR during a targeted
search of gamma-ray sources detected by \textit{Fermi} but not associated
with other known sources \citep{bassaetal17}.
Optical observations identify the binary companion of the pulsar,
and the companion star's low mass and short 6.42~hr orbital period
suggest \psr\ is in a black widow system, where the pulsar wind
irradiates and evaporates the companion star.
The position of \psr\ used for radio timing is that determined from its
optical counterpart, i.e., (R.A.,decl.[J2000])
$=(09^{\rm h}52^{\rm m}08.\!^{\rm s}319$,$-06^\circ 07\arcmin 23.\!\arcsec49$).
The distance is determined from the measured dispersion measure
($\DM=22.4\mbox{ pc cm$^{-3}$}$) and found to be either
$d=0.97\mbox{ kpc}$ or $1.74\mbox{ kpc}$, depending on which model of
Galactic electron distribution is used (NE2001 or YMW16, respectively).
Henceforth we assume a distance of 1.74~kpc, unless otherwise noted.
\cite{bassaetal17} estimate an interstellar absorption column
$\NH=4\times 10^{20}\mbox{ cm$^{-2}$}$
from a Galactic extinction model and distance,
which agrees with $\NH=3.9\times 10^{20}\mbox{ cm$^{-2}$}$ estimated from HI
in the direction of \psr\ \citep{dickeylockman90}.
We estimate a somewhat larger
$\NH=6.7_{-2.0}^{+2.9}\times 10^{20}\mbox{ cm$^{-2}$}$
(90\% confidence level) using the empirical relation between
$\NH$ and \DM\ from \citet{heetal13}.
Since \psr\ is in a short orbital period black widow system,
a higher $\NH$ could be measured,
e.g., as might be possible using dispersion measure variations
during radio eclipses such as that reported by \citet{mainetal18}
for PSR~B1957+20.
Using a short 4.6~ks exposure with \textit{Swift} XRT,
\citet{bassaetal17} obtain a $3\sigma$ upper limit on the 0.3--10 keV
flux $f_{0.3-10}<1.1\times 10^{-13}\mbox{ erg cm$^{-2}$ s$^{-1}$}$,
which corresponds to a X-ray luminosity limit of
$L<1.1\times 10^{31}\mbox{ erg s$^{-1}$}$ at 0.97~kpc or
$3.6\times 10^{31}\mbox{ erg s$^{-1}$}$ at 1.74~kpc.

In Section~\ref{sec:data}, we describe the \XMM\ observation of \psr\
and our procedure for processing the data.
Section~\ref{sec:spectra} and \ref{sec:variability} give details of
our spectral fitting and variability search analyses.
In Section~\ref{sec:discuss}, we summarize and discuss our results
and use them to place constraints on the r-mode instability.

\begin{figure}[ht]
\begin{center}
\includegraphics[width=0.45\textwidth,trim=0 48 0 0,clip]{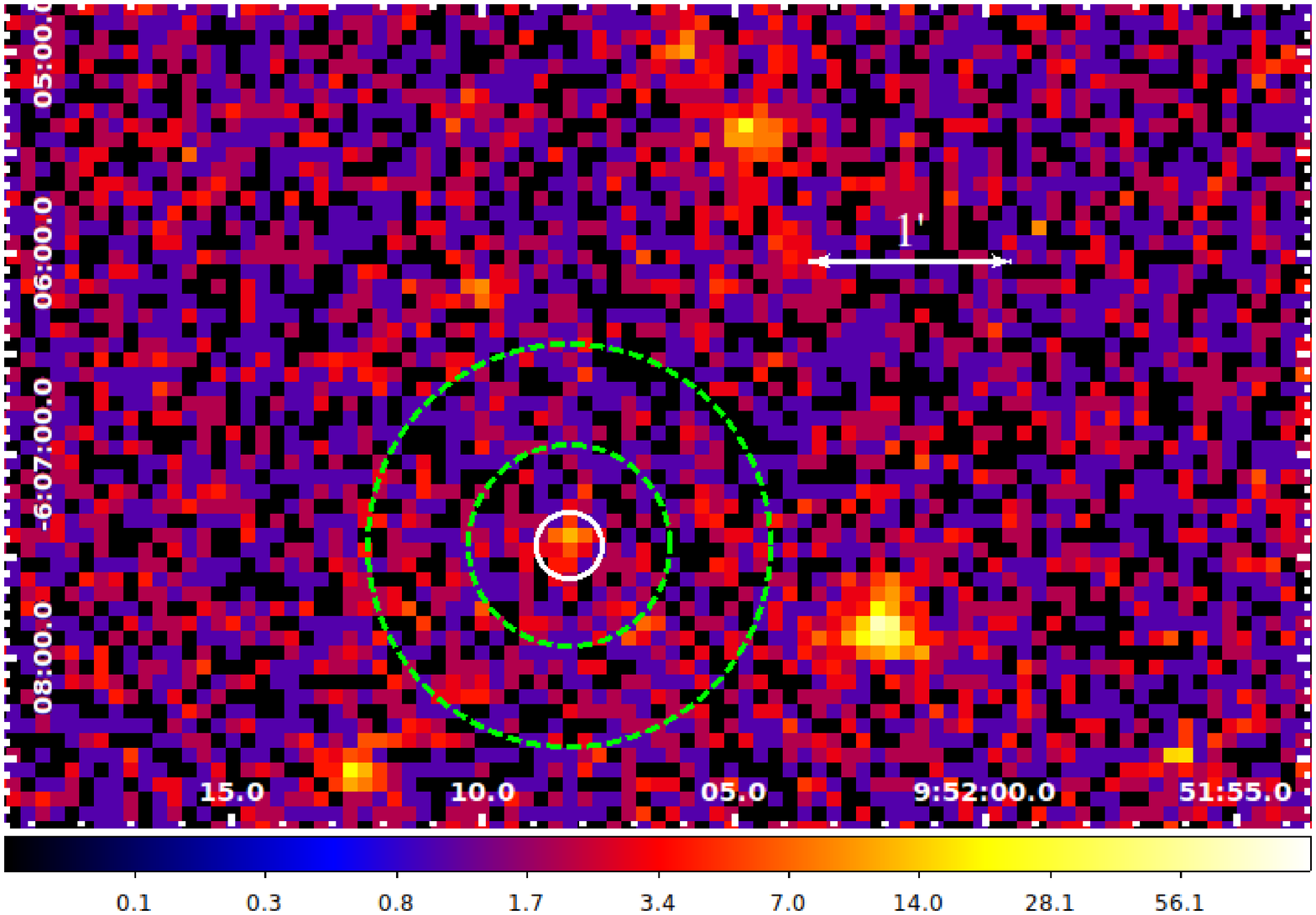}
\includegraphics[width=0.45\textwidth,trim=0 48 0 0,clip]{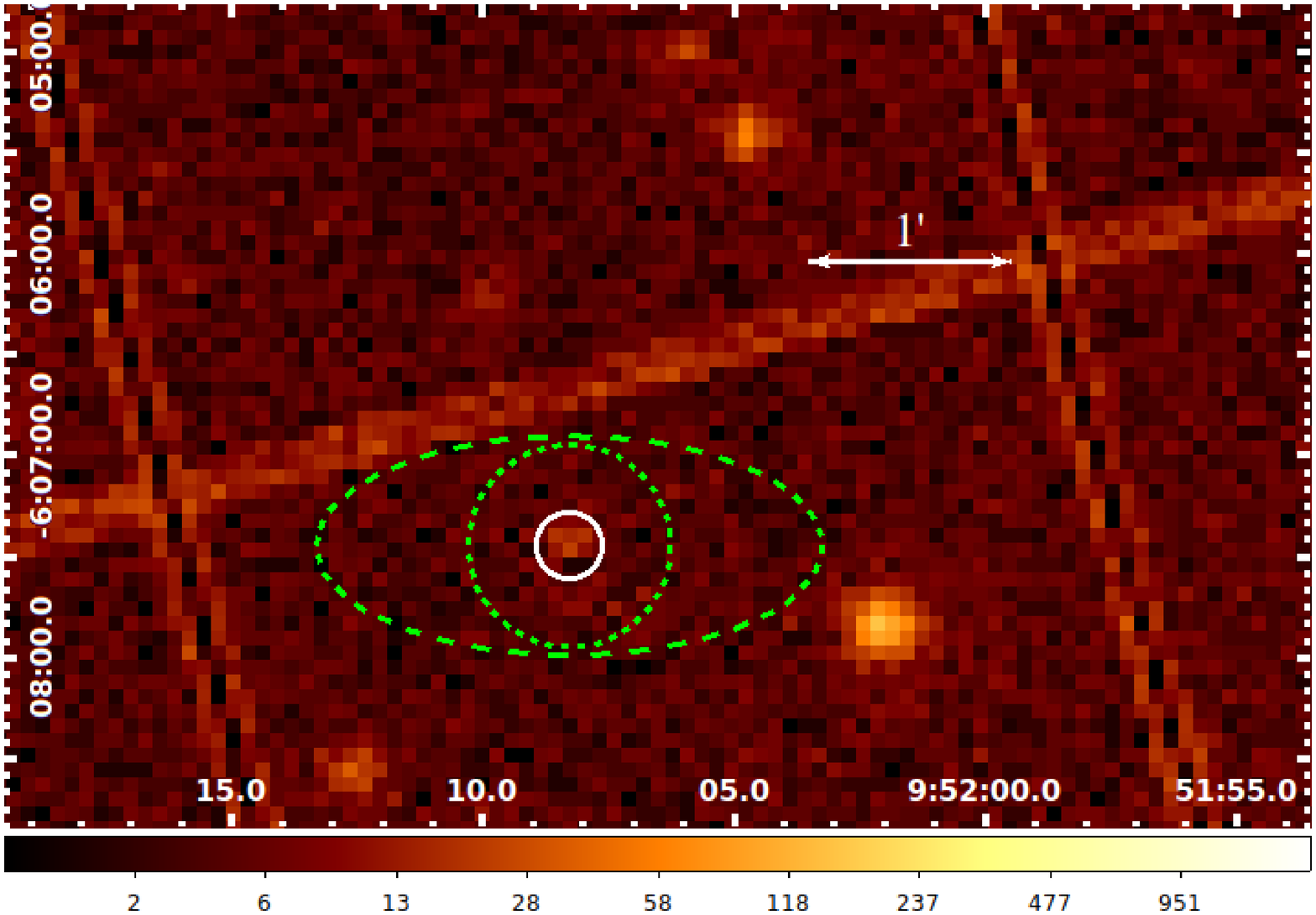}
\caption{MOS2 (top) and pn (bottom) images of the field of \psr.
North is up, and east is left.
In both images, the inner solid circle (of radius 10\arcsec) is used for
source spectral extraction.
In the MOS2 image, the dashed annulus (of inner and outer radii 30\arcsec\
and 60\arcsec, respectively) is used for background spectral extraction
(similarly for MOS1 data),
while in the pn image, the ellipsoidal dashed annulus (of circular inner radius
30\arcsec\ and maximum outer radius 75\arcsec) is used for background
spectral extraction.
\label{fig:fov}}
\end{center}
\end{figure}

\section{\XMM\ observations} \label{sec:data}

\XMM\ observed \psr\ on 2018 May 4 (ObsID 0821520101) for 71.2~ks with EPIC
in full frame imaging mode and using the thin optical filter.
Figure~\ref{fig:fov} shows the MOS2 and pn images of the field around \psr.
We process this data using SAS 17.0.0 \citep{gabrieletal04} and
CIAO 4.11 \citep{fruscioneetal06}.
Standard filtering is applied to both MOS and pn data, i.e.,
up to quadruple events are retained
and energy ranges of 0.2--12~keV for MOS and 0.2--15~keV for pn
are considered.
To remove periods of background flaring, we apply count-rate cuts of 2.1, 3,
and $18~\mbox{ s$^{-1}$}$ and obtain effective exposure times of 58.5, 63.3,
and 44.0~ks for MOS1, MOS2, and pn, respectively.
We then use \texttt{wavdetect} on the pn data to determine the source position
and its uncertainty to be (R.A.,decl.[J2000])
$=(09^{\rm h}52^{\rm m}08.\!^{\rm s}27\pm0.03$,
$-06^\circ 07\arcmin 27.\!\arcsec80\pm0.45$);
all errors are 1$\sigma$ unless otherwise noted.
With an angular resolution of $4.\!\arcsec1$ for pn
and estimated systematic uncertainty\footnote{\url{http://xmm2.esac.esa.int/docs/documents/CAL-TN-0018.pdf}}
in \XMM\ positions of $1.\!\arcsec2$ (1$\sigma$),
we can positively associate our X-ray point source with that of \psr.
We generate psf images of the MOS1, MOS2, and pn data using \texttt{psfgen},
centered on the source position at 0.5, 1, and 3~keV,
and do not find any noticeable extended emission.

We use \texttt{eregionanalyse} to determine optimum regions for
extracting MOS and pn source spectra.
With background regions shown in Figure~\ref{fig:fov},
this procedure indicated an optimal extraction radius of 10\arcsec\ and
background-subtracted counts of $39\pm10$, $65\pm12$, and $142\pm20$
and count-rates of $(0.67\pm0.18)\times10^{-3}$, $(1.02\pm0.19)\times10^{-3}$,
and $(3.24\pm0.44)\times10^{-3}\mbox{ s$^{-1}$}$ for MOS1, MOS2,
and pn, respectively.
Using \texttt{epatplot}, we find our spectra of \psr\ are not affected
by pile-up.
We account for bad pixels and chip gaps using \texttt{backscale}.
We then compute rmf and arf files.
In order to improve statistics, we combine the MOS1 and MOS2 spectra using
\texttt{epicspeccombine}.
Spectra are binned using ftools task \texttt{grppha} to a minimum of 15
photons per bin for each of the combined MOS spectrum and pn spectrum.

We perform spectral fitting using Xspec 12.10.1 \citep{arnaud96}.
We use \texttt{constant} to model a possible instrumental difference
between MOS and pn spectral normalizations, and we fix its value to 1
for the pn spectrum and allow it to vary for the combined MOS spectrum.
To model X-ray absorption by the interstellar medium, we use \texttt{tbabs}
with abundances from \citet{wilmsetal00}.
To model the intrinsic spectrum of \psr, we consider either a single
component composed of a power law (PL; \texttt{powerlaw}), blackbody
(BB; \texttt{bbodyrad}), or neutron star atmosphere,
or two components composed of combinations of the above.
For a (non-magnetic) neutron star hydrogen atmosphere model X-ray
spectrum, we use \texttt{nsatmos} \citep{heinkeetal06} and fix the model
parameters of neutron star mass and radius to $M=1.4\,\Msun$ and
$R=10\mbox{ km}$, respectively, and distance to $d=1.74\mbox{ kpc}$.
We also consider the non-magnetic atmosphere model \texttt{nsspec} with
an iron or solar composition, which are computed for fixed $M=1.4\,\Msun$
and $R=10\mbox{ km}$ and thus fixed surface gravity and gravitational
redshift \citep{gansickeetal02}.

\begin{deluxetable*}{cCCCCCC}[htb]
\tablecaption{Spectral model fits which allow for a varying $\NH$ \label{tab:spectra}}
\tablewidth{0pt}
\tablehead{
\colhead{Model fit parameter} & \colhead{PL} & \colhead{BB} & \colhead{BB+PL} & \colhead{NSATMOS+PL} & \colhead{NSATMOS+PL} & \colhead{2BB}
}
\startdata
$\NH$ ($10^{20}\mbox{ cm$^{-2}$}$) & 12.2_{-4.8}^{+7.2} & 2.6_{-2.2}^{+4.4} & 7.6_{-5.1}^{+11} & 11_{-7}^{+15} & 28.7_{-7.8}^{+8.3} & 5.2_{-3.6}^{+6.8} \\
$kT_\infty$ or $k\Teff$ (eV) & & 238_{-33}^{+36} & 175_{-50}^{+44} & 84_{-41}^{+44} & 38_{-4}^{+2} & 200_{-33}^{+35} \\
$R_\infty$ or $R_{\rm em}$ (m) \tablenotemark{a} & & 58_{-17}^{+25} & 110_{-46}^{+180} & 800_{-500}^{+5000} & {\rm fixed}\,10^4 & 84_{-28}^{+56} \\
$kT_\infty^{\rm hot}$ (eV) & & & & & & 1900_{-800}^{+2400} \\
$R_\infty^{\rm hot}$ (m) & & & & & & 1.3_{-0.2}^{+1.0} \\
$\Gamma$ & 2.51_{-0.39}^{+0.53} & & 0.78\pm0.85 & 0.59\pm0.90 & 1.38_{-0.61}^{+0.64} & \\
PL normalization ($10^{-6}$) & 1.74_{-0.35}^{+0.46} & & 0.34_{-0.25}^{+0.37} & 0.26_{-0.18}^{+0.56} & 0.76_{-0.39}^{+0.56} & \\
MOS-pn normalization & 1.23_{-0.23}^{+0.27} & 1.31_{-0.25}^{+0.31} & 1.25_{-0.22}^{+0.27} & 1.24_{-0.22}^{+0.27} & 1.22_{-0.22}^{+0.26} & 1.27_{-0.23}^{+0.28} \\
$f_{0.3-1}^{\rm abs}$ ($10^{-15}\mbox{ erg cm$^{-2}$ s$^{-1}$}$) & 1.9 & 1.9 & 2.0 & 2.0 & 2.1 & 2.0 \\
$f_{1-10}^{\rm abs}$ ($10^{-15}\mbox{ erg cm$^{-2}$ s$^{-1}$}$) & 3.6 & 1.3 & 7.6 & 8.0 & 6.3 & 7.1 \\
$f_{0.3-10}^{\rm abs}$ ($10^{-15}\mbox{ erg cm$^{-2}$ s$^{-1}$}$) & 5.5 & 3.2 & 9.6 & 10.0 & 8.3 & 9.1 \\
$\chi^2$/dof & 10.74/13 & 20.46/13 & 5.97/11 & 5.57/11 & 6.87/12 & 6.79/11 \\
\enddata
\tablenotetext{a}{Assuming $d=1.74\mbox{ kpc}$ to calculate $R_\infty$ and $R_\infty^{\rm hot}$ and $R=10\mbox{ km}$ to calculate $R_{\rm em}$.}
\tablecomments{Xspec models: PL=\texttt{powerlaw} and BB=\texttt{bbodyrad}.
For NSATMOS, $d=1.74\mbox{ kpc}$, $M=1.4\Msun$, and
$R=10\mbox{ km}$ are assumed.  All error bars are 1$\sigma$.
}
\vspace{-1.3em}
\end{deluxetable*}

\begin{figure}[htb]
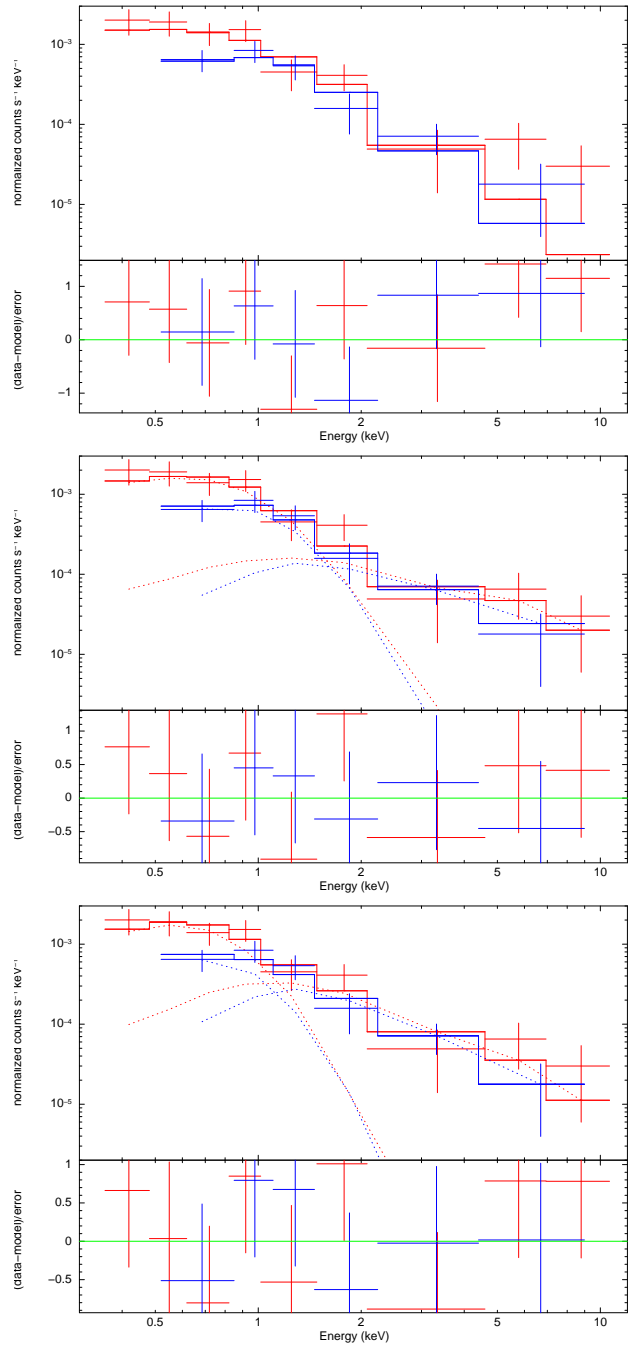

\begin{center}
\includegraphics[width=0.33\textwidth,angle=-90,trim=30 40 20 0,clip]{f2a.ps}
\includegraphics[width=0.33\textwidth,angle=-90,trim=30 40 20 0,clip]{f2b.ps}
\includegraphics[width=0.33\textwidth,angle=-90,trim=30 40 20 0,clip]{f2c.ps}
\caption{EPIC-pn (red) and combined MOS1 and MOS2 (blue) spectra
(upper panels) and model fit residuals (lower panels).
Solid lines are model fits, either \texttt{powerlaw} (top),
\texttt{nsatmos+powerlaw} (middle), and \texttt{nsatmos+powerlaw}
with fixed \texttt{nsatmos} normalization = 1 (bottom).
For two component models, dotted lines show individual components, i.e,
\texttt{nsatmos} at low energies and \texttt{powerlaw} at high energies.
\label{fig:spectra}}
\end{center}
\end{figure}

\section{Spectral analysis} \label{sec:spectra}

For our first set of spectral fits, we allow the absorption parameter $\NH$
to be free to vary.
The results of our simultaneous fit of the pn and combined MOS spectra are
given in Table~\ref{tab:spectra}.
The top panel of Figure~\ref{fig:spectra} shows the results of a
spectral fit using a \texttt{powerlaw} model.
A power law provides a generally good fit of the spectra of \psr, with a
photon index $\Gamma\approx 2.5_{-0.4}^{+0.5}$ and unabsorbed 0.3--10~keV flux
$f_{0.3-10}^{\rm unabs}\approx9\times 10^{-15}\mbox{ erg cm$^{-2}$ s$^{-1}$}$.
The latter results in a luminosity $L=3\times 10^{30}\mbox{ erg s$^{-1}$}$
at a distance of 1.74~kpc,
which is more than ten times lower than the upper limits determined
using a short \textit{Swift} exposure by \citet{bassaetal17}.
Even though a power law is a good fit to the data, there is possibly
unmodeled excess flux at the highest energies ($E\gtrsim 5\mbox{ keV}$),
the derived $\NH\approx(12_{-5}^{+7})\times 10^{20}\mbox{ cm$^{-2}$}$
is somewhat higher than the $(4-7)\times 10^{20}\mbox{ cm$^{-2}$}$
estimated to be in the direction of \psr\ (see Section~\ref{sec:intro}),
and the photon index $\Gamma\approx2.5$ is relatively soft and suggestive
that a thermal model is possibly more appropriate.
However, single component thermal models are a poor fit:
blackbody with $\chi^2/\mbox{dof}=20/13$ (see Table~\ref{tab:spectra}),
\texttt{nsatmos} with $\chi^2/\mbox{dof}=18/13$, and
\texttt{nsspec} with $\chi^2/\mbox{dof}=23/13$ for iron and
$\chi^2/\mbox{dof}=16/13$ for solar composition.
The poor fits are due to the model spectra not being able to match
the observed flux at high energies ($\gtrsim 2\mbox{ keV}$).
Note that, for the spectral fit using \texttt{nsatmos}, we allow the
model normalization to vary, and the fit results yield values smaller
than 1; while this is formally allowed and can be interpreted as
emission from only a fraction of the stellar surface, it is not strictly
correct since the model is computed assuming emission from the
entire surface.  If the normalization is fixed at one, then a much
larger distance would have to be assumed (for the same inferred flux
and temperature).

Two component spectral models produce improved fits because a thermal
component can reproduce the low energy spectrum and a power law or
second hot blackbody reproduces the high energy emission, even above 5~keV
(see Figure~\ref{fig:spectra}).
In the cases of BB+PL and 2BB, the derived $\NH$ (albeit with large
uncertainty) matches that inferred from the \DM\ of \psr.
However the resulting fit parameters are somewhat unusual
($\Gamma\sim0.8$ or $R_\infty^{\rm hot}=1\mbox{ m}$), and the two component
model fits are not strongly preferred
(f-test of BB+PL compared to PL yields a probability of 4.0\% of producing
by chance such a fit improvement when adding a blackbody).
A model fit using NSATMOS+PL with the normalization of \texttt{nsatmos}
free to vary (see above) gives results that are comparable to those of BB+PL.
On the other hand, if we fix this model normalization to be unity,
such that $R_{\rm em}=R=10\mbox{ km}$, then the surface temperature
$\Teff=4.4_{-1.0}^{+0.5}\times 10^5\mbox{ K}$ (at 90\% confidence),
and power law index $\Gamma\approx1.4$ is similar to that seen in other pulsars
(see, e.g., \citealt{delucaetal05,guillotetal16}, where $\Gamma=1.5-2.1$
is found for four pulsars), but
$\NH=(2.9\pm0.8)\times10^{21}\mbox{ cm$^{-2}$}$ is much greater that expected.
In this last case, a high \NH\ is needed to strongly reduce the low
energy model flux, which is high because the \texttt{nsatmos}
normalization is not allowed to be smaller and emission is from
the entire neutron star surface.
Similar results are obtained using \texttt{nsatmos+nsatmos},
where one component has a normalization fixed at one to model
emission from the entire surface while the other component with a
variable normalization could be that due to a hot spot.
Also problematic with this model fit is that the
temperature of the hot component hits the upper limit of the model
($\log\Teff=6.5$);
since the ratio of blackbody to atmosphere temperatures is seen to be
$\sim 2$ in spectral studies of other neutron stars, we expect this hot
component to have a temperature that exceeds $10^7\mbox{ K}$.
A f-test of NSATMOS+PL with fixed \texttt{nsatmos} normalization
compared to PL yields a probability of 2.3\%.

\begin{deluxetable*}{cCCCC}[htb]
\tablecaption{Spectral model fits with fixed $\NH$ and NSATMOS normalization (= 1) \label{tab:spectra_nh}}
\tablewidth{0pt}
\tablehead{
\colhead{Model fit parameter} & \colhead{PL} & \colhead{NSATMOS+PL} & \colhead{PL} & \colhead{NSATMOS+PL}
}
\startdata
fixed $\NH$ ($10^{20}\mbox{ cm$^{-2}$}$) & 4 & 4 & 10 & 10 \\
$k\Teff$ (eV) & & <18 & & <27 \\
$\Gamma$ & 1.85\pm0.15 & 1.85\pm0.14 & 2.36\pm0.19 & 2.36_{-0.25}^{+0.19} \\
PL normalization ($10^{-6}$) & 1.22\pm0.17 & 1.22\pm0.17 & 1.62\pm0.22 & 1.62_{-0.22}^{+0.20} \\
MOS-pn normalization & 1.27_{-0.24}^{+0.30} & 1.27_{-0.23}^{+0.29} & 1.23_{-0.23}^{+0.27} & 1.23_{-0.22}^{+0.27} \\
$f_{0.3-1}^{\rm abs}$ ($10^{-15}\mbox{ erg cm$^{-2}$ s$^{-1}$}$) & 1.6 & 1.6 & 1.9 & 1.9 \\
$f_{1-10}^{\rm abs}$ ($10^{-15}\mbox{ erg cm$^{-2}$ s$^{-1}$}$) & 5.3 & 5.3 & 3.9 & 3.9 \\
$f_{0.3-10}^{\rm abs}$ ($10^{-15}\mbox{ erg cm$^{-2}$ s$^{-1}$}$) & 6.9 & 6.9 & 5.8 & 5.8 \\
$\chi^2$/dof & 15.34/14 & 15.35/13 & 10.92/14 & 10.92/13 \\
\enddata
\tablecomments{Xspec model: PL=\texttt{powerlaw}.
For NSATMOS, $d=1.74\mbox{ kpc}$, $M=1.4\Msun$, and
$R=10\mbox{ km}$ are assumed.  All error bars are 1$\sigma$,
except for 90\% confidence level upper limit on $k\Teff$.
}
\vspace{-1.3em}
\end{deluxetable*}

In summary, while several models can fit well the spectrum of \psr,
none are entirely satisfactory because either the inferred $\NH$ is
too high or $\Gamma$ is too low or too high.
Nevertheless,
to obtain further constraints on the surface temperature of \psr,
we perform spectral fits using the NSATMOS+PL model with
\texttt{nsatmos} normalization fixed at 1 and absorption column fixed
at either $\NH=4\times 10^{20}$ or $1\times 10^{21}\mbox{ cm$^{-2}$}$,
i.e., values that span the likely range of \NH\ (see Section~\ref{sec:intro}).
At these low $\NH$ values compared to much higher values preferred when
$\NH$ is free to vary, the thermal component is forced to have a low
temperature (since the emission region is the entire stellar surface),
and the model fit is essentially that of a single component power law
(see Table~\ref{tab:spectra_nh}).
From these spectral fits, we derive upper limits (at 90\% confidence)
on the surface temperature of $2.1\times10^5$ and $3.1\times10^5\mbox{ K}$
for $\NH=4\times 10^{20}$ and $1\times 10^{21}\mbox{ cm$^{-2}$}$,
respectively.
Identical temperature limits are obtained when using non-magnetic,
fully-ionized helium atmosphere model spectra \citep{holai01}.
In order to obtain a surface temperature measurement, as opposed to an
upper limit, one needs $\NH\gtrsim 1.2\times 10^{21}\mbox{ cm$^{-2}$}$.

\section{Variability analysis} \label{sec:variability}

With a small inferred size of the X-ray emitting region, such as that
of a hot spot on the neutron star surface, there is the possibility of
detectable X-ray pulsations if the viewing geometry is favorable and
pulsations have a high enough amplitude.
Unfortunately, since the time resolution of our full frame imaging mode
observations is 2.6~s for MOS and 73.4~ms for pn, we are unable to search
for variability due to the pulsar spin period of 1.41~ms.
Nevertheless, we still perform an analysis to determine if there are
other periodic signals in the data. We consider MOS2 and pn independently.
We first apply a barycentric correction using \texttt{barycen} and DE405
ephemeris, then extract source and background 0.3--5~keV light curves
using the same spatial regions as in the spectral analysis, and obtain
corrected light curves using \texttt{epiclccorr}.
The resulting light curves of \psr\ with a bin size of 5000~s for
MOS2 data and 3000~s for pn data do not show clear evidence of
variability at the 6.42~hr orbital period;
note that our pn and MOS exposure times span 2 and 2.5 orbital cycles,
respectively.
Smaller bin sizes result in time intervals where the count-rate is zero.
We also do not find variability using \texttt{glvary}, with a variability
index of 0.

\section{Discussion} \label{sec:discuss}

In this work, we analyze a recent \XMM\ observation of the second
fastest pulsar known and report on our detection of its X-ray
counterpart.  The data are sufficient for extraction of source
spectra, and we find that these spectra can be fit well by a
single power law model or a two component thermal plus power law
model.  We do not detect any significant variability of the source,
although the \XMM\ full frame imaging observations do not allow us
to search for variations on the timescale of the short spin period
of \psr.

With gamma-ray flux
$f_\gamma=2.6\times10^{-12}\mbox{ erg cm$^{-2}$ s$^{-1}$}$,
spin period $P=1.41\mbox{ ms}$, and spin period time derivative
$\dot{P}=4.6\times 10^{-21}\mbox{ s s$^{-1}$}$ \citep{niederetal19},
\psr\ has $f_\gamma/f_{0.3-10}^{\rm unabs}\approx300$,
spin-down energy loss rate $\dot{E}=6.4\times10^{34}\mbox{ erg s$^{-1}$}$,
and $L/\dot{E}\approx 5\times10^{-5}$.
The gamma-ray to X-ray flux ratio is typical for black widow pulsars
(see, e.g., \citealt{marellietal15,salvettietal17}).
The $L/\dot{E}$ is well below $\sim 10^{-3}$ seen
in canonical rotation-powered pulsars \citep{becker09}
but similar to those seen in several millisecond pulsars in the Galactic
field (see, e.g., \citealt{kargaltsevetal12,leeetal18})
and globular clusters \citep{forestelletal14,bhattacharyaetal17}.
The spin-down rate of \psr\ could be sufficient to power its thermal
X-ray luminosity.
For example, deviations from beta equilibrium in the core
(via rotochemical heating) of \psr\ produce
a luminosity $\sim 5\times 10^{31}\mbox{ erg s$^{-1}$}$, but this
depends on uncertain properties of neutron superfluidity and proton
superconductivity \citep{reisenegger95,petrovichreisenegger10}.
Meanwhile, thermal creep of superfluid vortices dissipates energy at
a rate of $\sim (0.02-2)\times 10^{30}\mbox{ erg s$^{-1}$}$ in \psr\
\citep{alparetal84b,alparetal84},
and rotation-induced nuclear burning in the crust generates a heating
rate of $\sim 5\times 10^{29}\mbox{ erg s$^{-1}$}$ in \psr\
\citep{gusakovetal15}.

In Section~\ref{sec:spectra}, we use model fits of the \XMM\ spectrum
of \psr\ to constrain its surface temperature,
which in turn can be used to place limits on the neutron star core
temperature $\Tc$ through well-studied $\Teff$-$\Tc$ relations
(see, e.g., \citealt{potekhinetal15}).
For example, if the neutron star envelope is composed of iron, then
\begin{equation}
\Tc(\mbox{Fe}) = 1.29\times 10^8\mbox{ K}\left[
g_{14}^{-1}
\left(\frac{\Teff}{10^6\mbox{ K}}\right)^4
\right]^{5/11},
\label{eq:tstbfe}
\end{equation}
where $g\equiv10^{14}g_{14}\mbox{ cm s$^{-2}$}$ is surface gravity
\citep{gudmundssonetal82,potekhinetal97}.
On the other hand, for a fully accreted hydrogen envelope,
\begin{equation}
\Tc(\mbox{H}) = 0.552\times 10^8\mbox{ K}\left[
g_{14}^{-1}
\left(\frac{\Teff}{10^6\mbox{ K}}\right)^4
\right]^{7/17}
\label{eq:tstbh}
\end{equation}
\citep{potekhinetal97}.
Using equations~(\ref{eq:tstbfe}) and (\ref{eq:tstbh}), our
surface temperature measurement of $\Teff\approx 4.4\times 10^5\mbox{ K}$
(for varying $\NH$ and thermal plus power law model spectrum)
yields a core temperature of $\Tc=1.9\times 10^7\mbox{ K}$ for an
iron envelope or $\Tc=1.0\times 10^7\mbox{ K}$ for a hydrogen envelope.
Meanwhile, our surface temperature upper limit of
$\Teff<3.1\times 10^5\mbox{ K}$ (for fixed
$\NH=1\times10^{21}\mbox{ cm$^{-2}$}$ and simple power law model spectrum)
yields core temperatures of $\Tc<1.0\times 10^7\mbox{ K}$ for an iron
envelope or $\Tc<0.56\times 10^7\mbox{ K}$ for a hydrogen envelope.

\begin{figure}[tb]
\begin{center}
\includegraphics[width=0.47\textwidth]{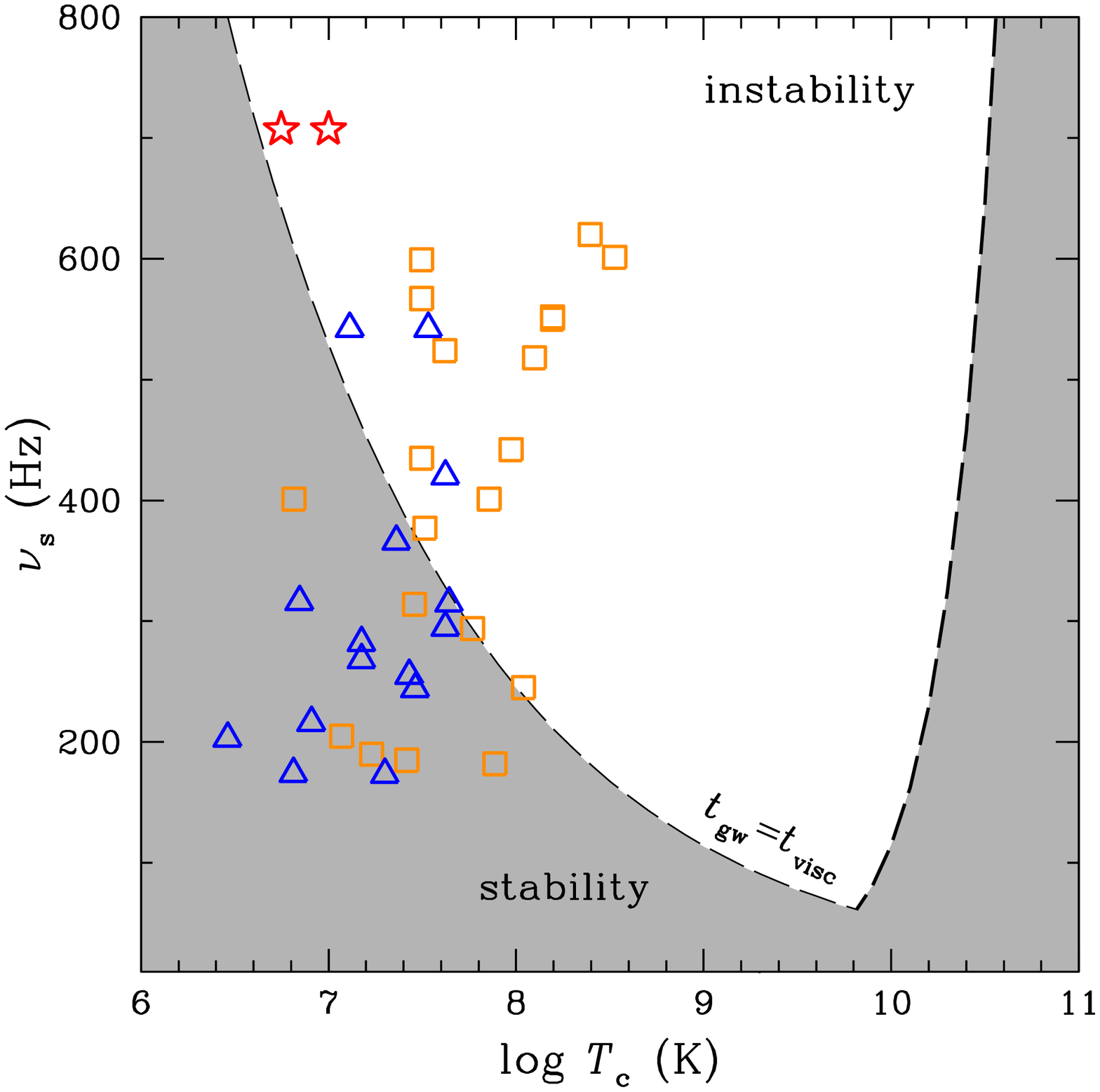}
\caption{R-mode instability window.
Neutron star core temperature $\Tc$ versus spin frequency $\nus$.
Dashed line denotes the boundary set by $\tgw=\tvisc$, with the shaded
``stability'' region set by $\tgw>\tvisc$ and r-mode is damped by
viscosity and unshaded ``instability'' region set by $\tgw<\tvisc$ and
r-mode grows by emission of gravitational waves.
Stars indicate ($\Tc$,$\nus$) for \psr, where the higher $\Tc$ is
obtained from spectral fitting with variable $\NH$ and lower $\Tc$ from
fixed $\NH=1\times10^{21}\mbox{ cm$^{-2}$}$ (see text).
Triangles and squares denote millisecond pulsars and neutron stars in a
low-mass X-ray binary, respectively (see text).
\label{fig:rmode}}
\end{center}
\end{figure}

As noted in Section~\ref{sec:intro}, the r-mode instability is caused
by growth of the oscillation mode through emission of gravitational waves
and occurs on a timescale
\begin{equation}
\tgw = 47\mbox{ s}
\left(\frac{1000\mbox{ Hz}}{\nus}\right)^6, \label{eq:tgw}
\end{equation}
while the oscillation is damped by viscosity on a timescale $\tvisc$,
which in the simplest model is due to
electron-electron scattering (once the core temperature drops below
the neutron superfluid transition temperature) with
\begin{equation}
\tvisc = 2.2\times 10^5\mbox{ s}
\left(\frac{\Tc}{10^8\mbox{ K}}\right)^2 \label{eq:tee}
\end{equation}
\citep{anderssonkokkotas01,shterninyakovlev08,gusakovetal14}.
Therefore the r-mode is unstable and grows when $\tgw<\tvisc$ and is
damped when $\tgw>\tvisc$.
This is illustrated in Figure~\ref{fig:rmode}, which displays the
``r-mode instability window'' based on the two primary parameters,
$\nus$ and $\Tc$, that determine $\tgw$ and $\tvisc$;
note that the dominant viscosity at $\Tc\gtrsim 10^{10}\mbox{ K}$ is not
that given by equation~(\ref{eq:tee}) but by bulk viscosity.
Also shown are ($\nus$,$\Tc$) for neutron stars in systems where
these values or their limits can be determined (see
\citealt{gusakovetal14,bhattacharyaetal17,mahmoodifarstrohmayer17,rangelovetal17,schwenzeretal17,gonzalezcaniulefetal19},
and references therein; see also \citealt{mahmoodifarstrohmayer13};
for simplicity, we plot fiducial $\Tc$ from \citealt{gusakovetal14},
but we note there is factor of $\lesssim 3$ uncertainty due to uncertain
envelope composition and gravitational redshift).
For \psr, we show two values of $\Tc$ derived from our spectral fits
using either a varying or fixed $\NH$ and assuming a fully accreted
hydrogen envelope.

\begin{figure*}[htb]
\begin{center}
\includegraphics[width=0.47\textwidth]{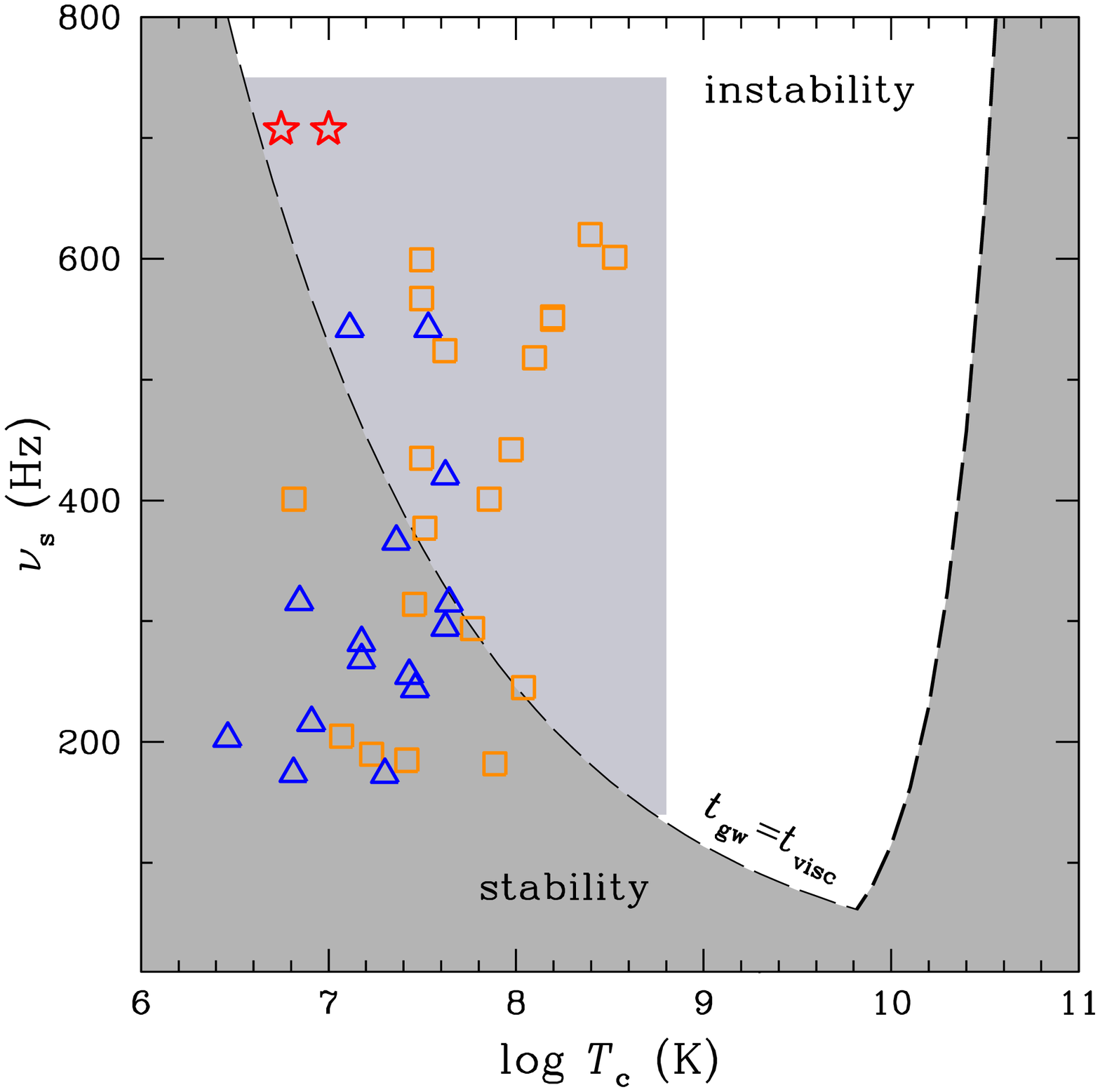}
\hspace{0.8cm}
\includegraphics[width=0.47\textwidth]{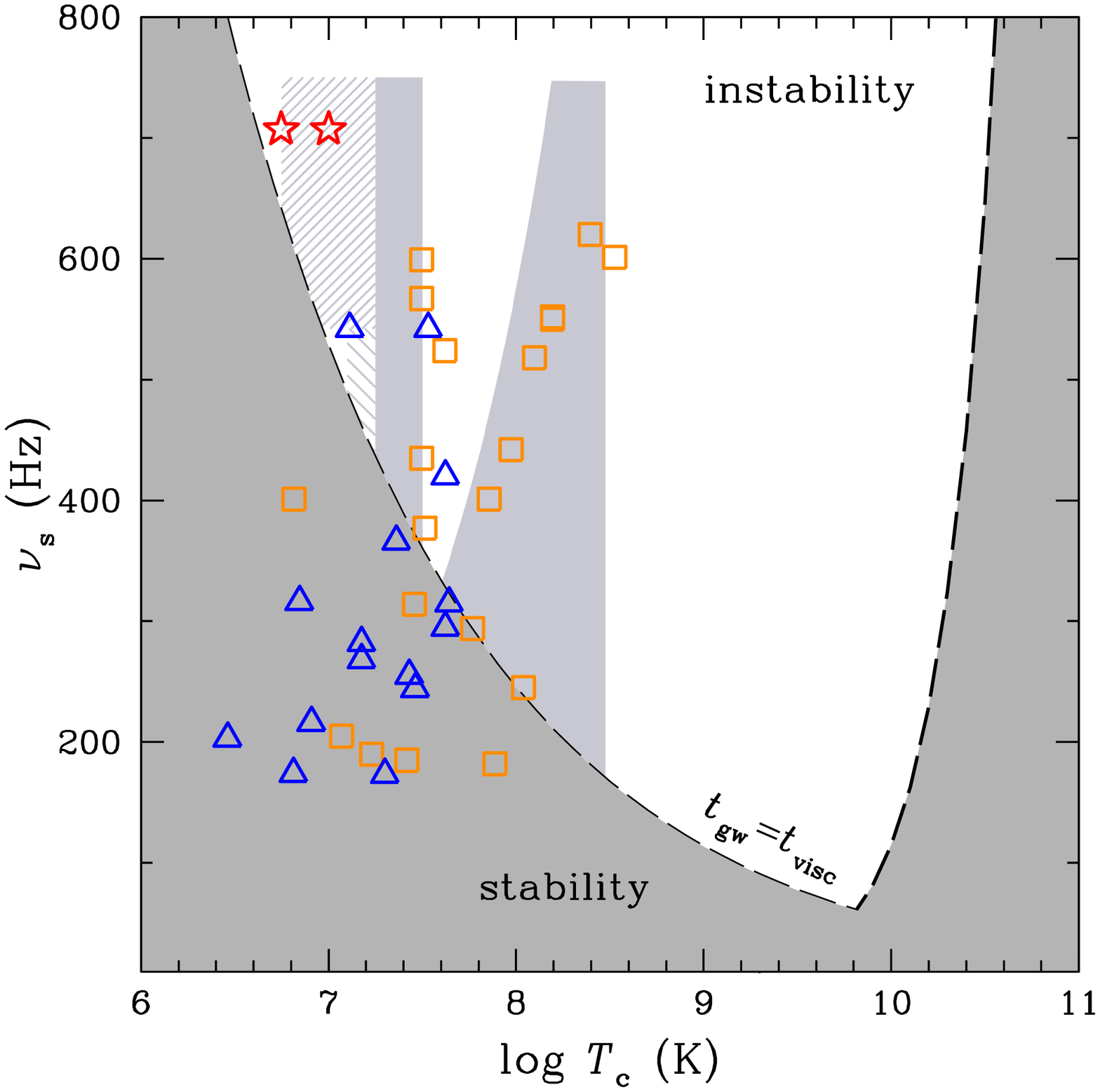}
\caption{R-mode instability window.
Neutron star core temperature $\Tc$ versus spin frequency $\nus$.
Dark shading denotes stability region due to shear viscosity by
electron-electron scattering, and no shading denotes instability region.
($\Tc$,$\nus$) for \psr, millisecond pulsars, and neutron stars in
a low-mass X-ray binary are denoted by stars, triangles, and squares,
respectively.  See Figure~\ref{fig:rmode} caption for more details.
Left panel: Light shading denotes stability region due to superfluid mutual
friction.
Right panel: Light shading and hatching denote regions required to stabilize
r-modes in observed systems.
\label{fig:rmode2}}
\end{center}
\end{figure*}

Our results for \psr\ can be used to inform our understanding of r-modes.
Figure~\ref{fig:rmode} shows that many neutron stars, including \psr\
(see below),
should be unstable to r-mode growth and thus are potentially strong
sources of gravitational waves.  This in part motivates r-mode searches
by LIGO/Virgo \citep{abbottetal17,meadorsetal17,abbottetal19,carideetal19}.
However, spin and thermal evolution calculations indicate neutron stars
should only spend a short time within the instability window and long
time outside the window \citep{levin99,heyl02}.
Thus one expects few sources to lie within the window at any one
time compared to the number of sources outside the window,
contrary to what is shown in Figure~\ref{fig:rmode}.
This suggests that the instability window should be much smaller and
our theoretical understanding of the physics that determines $\tvisc$
(e.g., neutrino emission, crust elasticity, superfluidity, and/or
existence of hyperonic or quark matter;
\citealt{anderssonkokkotas01,alfordetal12,alfordetal12b}) is inadequate
\citep{hoetal11,haskelletal12,chugunovetal17}.
For example, one possible mechanism that could close the window at
$\Tc<10^9\mbox{ K}$ is superfluid mutual friction,
i.e., dissipation due to interactions between protons/electrons and
superfluid neutron vortices \citep{haskelletal09}.
Its possible effect on the r-mode instability window is illustrated
in the left panel of Figure~\ref{fig:rmode2}
(see \citealt{haskelletal09,haskelletal12}, for details;
see also \citealt{hoetal11}).

As noted above, \psr\ is inside the traditional r-mode instability
region shown in Figure~\ref{fig:rmode}.
If the measured rate of change of spin frequency $\nudot$ is due
(entirely) to energy loss by gravitational wave emission at a constant
r-mode oscillation amplitude $\alpha$, then the amplitude is
\begin{equation}
\alpha \approx 2\times10^{-8}
\left(\frac{1000\mbox{ Hz}}{\nus}\right)^{7/2}
\left(\frac{\nudot}{10^{-15}\mbox{ Hz s$^{-1}$}}\right)^{1/2}
\label{eq:alphas}
\end{equation}
\citep{owenetal98,anderssonkokkotas01}
and is $\alpha=1\times10^{-7}$ for \psr.
Another constraint on $\alpha$ is set by the balance between viscous
heating (by the aforementioned electron-electron scattering) of an
unstable r-mode and cooling by surface radiation (at these low
temperatures, surface photon emission dominates core neutrino emission).
This yields
\begin{eqnarray}
\alpha &=& 1.1\times10^{-9}
\left(\frac{L_{\rm th}}{10^{30}\mbox{ erg s$^{-1}$}}\right)^{1/2}
\left(\frac{\Tc}{10^7\mbox{ K}}\right)
\left(\frac{1000\mbox{ Hz}}{\nus}\right) \nonumber\\
&\sim& 3\times10^{-10}
\left(\frac{L_{\rm th}}{10^{30}\mbox{ erg s$^{-1}$}}\right)^{31/34}
\left(\frac{1000\mbox{ Hz}}{\nus}\right)
 \label{eq:alphaheat}
\end{eqnarray}
\citep{owenetal98,anderssonkokkotas01},
where $L_{\rm th}$ is thermal luminosity and the second equality makes
use of equation~(\ref{eq:tstbh}).
Alternatively, if we assume that \psr\ lies on the boundary between
stability and instability, such that $\tgw=\tvisc$, then balance of
heating and cooling yields
\begin{equation}
\alpha = 1.6\times10^{-10}
\left(\frac{L_{\rm th}}{10^{30}\mbox{ erg s$^{-1}$}}\right)^{1/2}
\left(\frac{1000\mbox{ Hz}}{\nus}\right)^4
 \label{eq:alphaheat2}
\end{equation}
\citep{mahmoodifarstrohmayer13}.
Since we see from Figure~\ref{fig:rmode} that \psr\ is near the
boundary, such that $\tgw\sim\tvisc$, equations~(\ref{eq:alphaheat}) and
(\ref{eq:alphaheat2}) give similar values of $\alpha\sim1\times10^{-9}$.
These constraints on r-mode amplitude are among the strongest obtained
thus far.  For example, X-ray observations yield upper limits of
$\alpha\approx 10^{-8}-10^{-6}$ for millisecond pulsars and
neutron stars in a low-mass X-ray binary
\citep{mahmoodifarstrohmayer13,chugunovetal17,mahmoodifarstrohmayer17,schwenzeretal17}
and of $\sim 2\times10^{-9}$ for the 542~Hz pulsar 47~Tucanae~aa
\citep{bhattacharyaetal17}.

A phenomenological approach to constraining the shape of the r-mode
instability window with observed millisecond pulsars and low mass
X-ray binaries is suggested by \citet{chugunovetal17}.
They find observed systems require suppression of the instability in
two extra regions: at $\Tc\sim 10^8$~K and at
$2\times 10^7\mbox{ K}\lesssim\Tc\lesssim 3\times 10^7\mbox{ K}$
(see light shaded regions in right panel of Figure~\ref{fig:rmode2}).
The first region is required for consistency with the
hottest neutron stars in a low mass X-ray binary, while the second
is needed for colder neutron stars in a low mass X-ray binary and
upper limits on the surface temperature of millisecond pulsars.
Extension of the second constraint to even lower temperatures
(see hatched regions in right panel of Figure~\ref{fig:rmode2})
may be needed due to surface temperature limits of 47~Tucanae~aa
\citep{bhattacharyaetal17} and \psr\ presented here,
such that the r-mode instability is not active at
$\Tc\lesssim 3\times 10^7\mbox{ K}$,
although isolated regions of instability are not excluded
(see, e.g., \citealt{gusakovetal14}).
Finally, \citet{chugunovetal17} argue that the r-mode instability
can be almost unsuppressed at
$7\times 10^7\mbox{ K}\lesssim\Tc\lesssim 10^8\mbox{ K}$
while still remaining consistent with observations;
in such a case, this leads to a class of rapidly rotating non-accreting
neutron stars which are heated by the r-mode instability
(see also \citealt{chugunovetal14}).

Gravitational waves from \psr\ could be detectable if the r-mode
amplitude $\alpha\gtrsim 1\times10^{-7}(h_0/h_{\rm sd})$ or the pulsar
has an ellipticity $\varepsilon\gtrsim 6.9\times10^{-10}(h_0/h_{\rm sd})$,
where $h_0$ is gravitational wave strain amplitude and $h_{\rm sd}$
is the ``spin-down limit'' amplitude obtained by assuming the pulsar's
entire rotational energy loss is due to gravitational wave emission;
note that $h_0$ and $h_{\rm sd}$ depend on frequency and gravitational
wave mechanism
(e.g., $h_{\rm sd}\approx 8\times10^{-28}$ for an ellipticity in \psr).
Gravitational wave searches of \psr\ using LIGO 2015--2017 data are
sensitive to $h_0/h_{\rm sd}\sim 60$ and thus are not able to
physically constrain $\varepsilon$ \citep{abbottetal19b,niederetal19}.

Finally, the observing mode of the \XMM\ data presented here is not able to
resolve pulsations at the 1.41~ms spin period of \psr.
Detection of pulsations using \XMM\ timing mode (with a time resolution
of 0.03~ms) could permit discrimination between spectral models and
determination of the size of the X-ray emitting region on the neutron star.
Modeling of the pulse profile could even lead to some constraints on
the nuclear equation of state (e.g., \citealt{bogdanov13}), such as
that being done with observations of other millisecond pulsars using
\textit{NICER} \citep{gendreauetal12}.

\acknowledgments

The authors thank the anonymous referee for useful comments.
WCGH appreciates use of computer facilities at the Kavli Institute for
Particle Astrophysics and Cosmology.
WCGH acknowledges support through grants 80NSSC19K0668 from NASA and
ST/R00045X/1 from the Science and Technology Facilities Council in
the United Kingdom.
COH is supported by NSERC Discovery Grant RGPIN-2016-04602 and a Discovery
Accelerator Supplement.
AIC acknowledges support by RFBR according to the research project 18-32-20170.
This work is based on observations obtained with \XMM, an ESA science
mission with instruments and contributions directly funded by ESA Member
States and NASA.

\vspace{5mm}
\facility{\XMM}

\software{CIAO 4.11 \citep{fruscioneetal06}, SAS 17.0.0 \citep{gabrieletal04},
 Xspec 12.10.1 \citep{arnaud96}}

\bibliography{psrj0952v2.bib}

\end{document}